\documentclass[modern]{aastex62}
\usepackage{longtable}

\usepackage{ulem}

\usepackage{graphicx}

\hypersetup{linkcolor=cyan,citecolor=cyan,filecolor=cyan,urlcolor=cyan}
\usepackage{amsmath}

\received{January 1, 2018}
\revised{January 7, 2018}
\accepted{\today}
\submitjournal{ApJ}

\shorttitle{The changing rotational light curve amplitude of Varuna}
\shortauthors{Fern\'andez-Valenzuela et al.}

\begin{document}

\title{The changing rotational light-curve amplitude of Varuna and evidence for a close-in satellite}

\correspondingauthor{Estela Fern\'andez-Valenzuela}
\email{estela@ucf.edu, ortiz@iaa.es}

\author[0000-0002-0786-7307]{Estela Fern\'andez-Valenzuela}
\affil{Florida Space Institute\\
12354 Research Parkway, Partnership 1, Room 211 \\
Orlando, FL 32826-0650, USA}

\author{Jose Luis Ortiz}
\affiliation{Instituto de Astrof\'isica de Andaluc\'ia (IAA-CSIC) \\
Glorieta de la Astronom\'ia s/n \\
Granada, 18008, Spain}

\author{Nicol\'as Morales}
\affiliation{Instituto de Astrof\'isica de Andaluc\'ia (IAA-CSIC) \\
Glorieta de la Astronom\'ia s/n \\
Granada, 18008, Spain}

\author{Pablo Santos-Sanz}
\affiliation{Instituto de Astrof\'isica de Andaluc\'ia (IAA-CSIC) \\
Glorieta de la Astronom\'ia s/n \\
Granada, 18008, Spain}

\author{Ren\'e Duffard}
\affiliation{Instituto de Astrof\'isica de Andaluc\'ia (IAA-CSIC) \\
Glorieta de la Astronom\'ia s/n \\
Granada, 18008, Spain}

\author{Amadeo Aznar}
\affiliation{Observatorio Isaac Aznar, Grupo de Observatorios APT \\
C/La Plana, 44, 13 \\
E-46530 Pu\c{c}ol, Valencia, Spain}

\author{Vania Lorenzi}
\affiliation{Fundaci\'on Galileo Galilei -- Istituto Nazionale di Astrofisica \\
Rambla Jos\'e Ana Fern\'andez P\'erez, 7\\
38712 Bre\~na Baja, TF, Spain}
\affiliation{Instituto de Astrof\'isica de Canarias\\
 C/Va L\'actea s/n\\
 38205 La Laguna, Spain}

\author{Noem\'i Pinilla-Alonso}
\affiliation{Florida Space Institute\\
12354 Research Parkway, Partnership 1, Room 211 \\
Orlando, FL 32826-0650, USA}

\author{Emmanuel Lellouch}
\affiliation{LESIA, Observatoire de Paris, Universit\'e PSL, CNRS, Univ. Paris Diderot, Sorbonne Paris\\
1Cit\'e, Sorbonne Universit\'e, 5 Place J. Janssen \\
 92195 Meudon Pricipal Cedex, France} 


\begin{abstract}
From CCD observations carried out with different telescopes, we present short-term photometric measurements of the large trans-Neptunian object Varuna in 10 epochs, spanning around 19 years. We observe that the amplitude of the rotational light-curve has changed considerably during this period of time from 0.41 to 0.55 mag. In order to explain this variation, we constructed a model in which Varuna has a simple triaxial shape, assuming that the main effect comes from the change of the aspect angle as seen from Earth, due to Varuna's orbital motion in the 19-year time span. The best fits to the data correspond to a family of solutions with axial ratios $b/a$ between 0.56 and 0.60. This constrains the pole orientation in two different ranges of solutions presented here as maps. Apart from the remarkable variation of the amplitude, we have detected changes in the overall shape of the rotational light-curve over shorter time scales. After the analysis of the periodogram of the residuals to a 6.343572~h double-peaked rotational light-curve fit, we find a clear additional periodicity. We propose that these changes in the rotational light-curve shape are due to a large and close-in satellite whose rotation induces the additional periodicity. The peak-to-valley amplitude of this oscillation is in the order of 0.04 mag. We estimate that the satellite orbits Varuna with a period of 11.9819 h (or 23.9638 h), assuming that the satellite is tidally locked, at a distance of $\sim1300$ km (or $\sim2000$ km) from Varuna, outside the Roche limit.
\end{abstract}



\section{Introduction} \label{sec:intro}

Trans-Neptunian objects (TNOs) are solar system bodies that orbit the Sun with larger semi-major axis than that of Neptune, formed quite outside the so-called ``snow line'' where the temperature of the proto-planetary disk was low enough to allow the survival of molecules of chemical compounds with low sublimation points. Due to the ample distances that separate TNOs from the Sun, they have suffered less chemical processes than other solar system bodies and may preserve material from the nebula that generated the solar system. Hence, TNOs yield important information about the solar system formation and its evolution.

Currently, observational searches have found $\sim\ 2500$ TNOs (according to the data from MPC\footnote{\url{https://minorplanetcenter.net}}), although dynamical models suggest that the TNO population located between $30-50$ au from the Sun is around $10^5$ objects with diameters larger than 100 km \citep{Trujillo2001,Petit2008,Petit2011}.

20000 Varuna is one of the most interesting TNOs due to its peculiar physical properties. It rotates relatively fast with a period of $6.34358\pm0.00005$~h \citep{Belskaya2006,Santos-Sanz2013}, producing a double-peaked rotational light-curve dominated by its shape \citep[e.g.,][]{Jewitt2002,Lellouch2002,Hicks2005,Belskaya2006}. The rotational light-curve amplitude reported in these studies is quite large, $\sim0.45$ mag, indicating that the body is highly elongated (i.e., $b/a<0.66$, being $a$ and $b$ the larger semi-axes of a triaxial body). The fast period and the elongated shape require a density of $\sim1000$ kg$\,$m$^{-3}$ \citep[assuming hydrostatic equilibrium;][]{Chandrasekhar1987}. This is somewhat high compared to other TNOs of similar sizes, considering that Varuna's equivalent diameter is $\sim700$ km,  given by thermo-physical models using Herschel Space Telescope measurements \citep{Lellouch2013}; see the supplementary material in \cite{Ortiz2012} to compare the density of objects with similar sizes. However, a stellar occultation by Varuna, detected in 2010, results in a long chord of 1004 km \citep{Sicardy2010}. This value is somewhat in tension with the one given by \cite{Lellouch2013}, but fits better with Varuna's density.

Motivated by all of this, we have been monitoring Varuna for nearly two decades now, resulting in a collection of data with several rotational light-curves (light curves in the following). In this work, we present new observations of Varuna at different epochs (section \ref{sec:observations}). The results from these observations, together with other observations from the literature, are presented in section \ref{sec:results}. Section \ref{sec:analysis} contains the analysis of the set of light curves in order to obtain Varuna's pole orientation and shape. Section \ref{sec:satellite} displays the evidence for a possible satellite. A brief summary is presented in section \ref{sec:conclusions}.

\section{Observations and data reduction} \label{sec:observations}

We carried out eight observing campaigns from 2005 to 2019 using four different telescopes. A log of observations is shown in online table 1.

The 1.5-m telescope at Sierra Nevada Observatory in Granada (Spain) has a CCD VersArray with $2048\times2048$ pixels, while the field of view (FoV) is $7.92'\times7.92'$ and the image scale is $0.232''/$pixel. Observations were obtained in $2\times2$ binning mode. The typical seeing was $\sim1.5''$, being the point spread function well sampled for the photometric goals.

The 1.23-m and the 2.2-m telescopes at Calar Alto Observatory are located in Almer\'ia (Spain). The 1.23-m telescope uses a DLR-MKIII camera with $4096\times4096$ pixels, while the FoV and the image scale are $21.5'\times21.5'$ and  $0.315''/$pixel, respectively. At the 2.2-m telescope we used two different instruments: the Bonn University Simultaneous CAmera (BUSCA) and the Calar Alto Faint Object Spectrograph (CAFOS). BUSCA allows the simultaneous direct imaging of the same sky area in four colors; each CCD has $4096\times4096$ pixels, with a FoV of $12'\times12'$ and an image scale of $0.176''/$pixel. We used $2\times2$ binning mode. CAFOS is equipped with a $2048\times2048$ pixel CCD; we used the SITe\#1d chip which produces a $0.53''/$pixel image scale and a circular FoV of $16'$.

The Telescopio Nazionale Galileo (TNG), with 3.58-m diameter, is located at the Roque de los Muchachos Observatory in Canary Islands (Spain). We used the Device Optimized for the LOw RESolution (DOLORES) with a detector of $2048\times2048$ pixels. The image scale is $0.252''/$pixel which yields a FoV of $8.6'\times8.6'$. 

Bias and twilight flat-field calibration images were taken at the beginning of each observation night. The science images were processed by subtracting a median bias and dividing by a median flat-field. Using the Interactive Data Language (IDL), we developed our own code to perform these reduction steps, as well as the synthetic aperture photometry for the time series data.

We performed the synthetic aperture photometry using different aperture radii, as described in \cite{FernandezValenzuela2016}, and choosing the one in which the dispersion of the photometry is minimal. To construct the light curves, Varuna was compared to a set of reference stars located as close as possible to the object. Whenever possible during each campaign, we employed the same set of comparison stars each night in order to minimize systematic photometric errors.

\section{Results from observations}
\label{sec:results}
\begin{figure}
\plotone{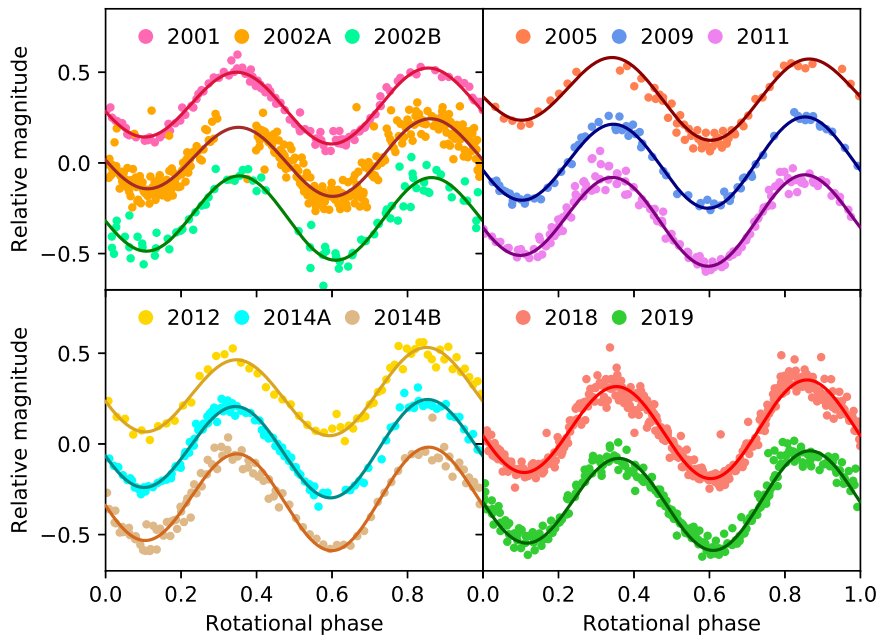}
\caption{Varuna's light curves from 2001 to 2019. Each color corresponds to a different epoch. 2001, 2002A and 2002B data were taken from \cite{Jewitt2002,Lellouch2002} and \cite{Hicks2005}, respectively. The lines over plotted to the data points represent the fit of the observational data to the Fourier series. Error bars are not shown to avoid a cluttered plot; the data with the uncertainties are given in the online table 2. \label{fig:rotational_light_curves}}
\end{figure}

From the time series observations obtained between 2005 and 2019, we have built eight different light curves. Additionally, we used data published in the literature by \cite{Jewitt2002,Lellouch2002} and \cite{Hicks2005}, thus incorporating three additional light curves to our study from previous years (figure \ref{fig:rotational_light_curves}). All the light curves were corrected from light travel time. Because Varuna's body is assumed to have an ellipsoidal shape \cite[e.g.,][]{Jewitt2002,Lellouch2002}, data from each light curve were fitted to a Fourier series $m=\Sigma_{i}\left[a_i\sin(2i\pi\phi)+b_i\cos(2i\pi\phi)\right]$, where $m$ is the theoretical value of the relative magnitude obtained from the fit, $\phi$ is the rotational phase (calculated as the fractional part of $({\rm JD}-{\rm JD_0})/P$, where JD is the Julian Date, JD$_0=2451957.0$ is the initial Julian Date, and $P$ is the rotation period in days), and $(a_i$, $b_i)$ are the coefficients of the Fourier function (with $i=0,1,2,...$). In our specific case, we used up to second-order ($i=2$) or up to fourth-order ($i=4$) Fourier functions. The second order is the minimum order that allows a double-peaked fit; however, higher orders take into account small deviations on inhomogeneous objects and can be used to fit light curves that are highly sampled. The 2001, 2002A, 2011, 2018 and 2019 light curves were fitted to a fourth-order function, while the remaining light curves were fitted to a second-order Fourier function, because the number of data points in those runs was not large enough to use a higher order. Data were folded using Varuna's rotation period of $6.343572\pm0.000006$~h, which is obtained using the Lomb periodogram analysis of all our data, in agreement with the previous one reported in \cite{Belskaya2006}. The peak-to-valley amplitudes $\Delta m$ (amplitudes in the following) of each light curve are given by the absolute maximum and minimum produced by the fits. 

Table \ref{tb:amplitudes} contains the results from the fit to each light curve, i.e., the amplitude and the dispersion of the residuals of the Fourier function fit to the observational data. One of the evident results is that the amplitude has changed considerably along these 19 years, with an increase of $\sim0.13$ mag. Online table 2 presents all the relative photometry observations from 2005 to 2019.

\begin{deluxetable}{lcc}
\tablecaption{Varuna's light curve amplitudes from different epochs. Abbreviations are defined as follows: light curve amplitude ($\Delta m$); dispersion of the residuals' values of the second-order Fourier function fit to the observational data ($\sigma$). \label{tb:amplitudes}}
\tablecolumns{3}
\tablenum{1}
\tablewidth{0pt}
\tablehead{
\colhead{Year} &
\colhead{$\Delta m$} &
\colhead{$\sigma$}  \\
\colhead{} & \colhead{(mag)} &
\colhead{(mag)}
}
\startdata
2001\tablenotemark{a} & $0.41\pm0.01$ & 0.037 \\
2002A\tablenotemark{b} & $0.45\pm0.01$ & 0.057  \\
2002B\tablenotemark{c} & $0.47\pm0.02$ & 0.064  \\
2005 & $0.46\pm0.02$ &  0.036  \\
2009 & $0.50\pm0.01$ &  0.024  \\
2011 & $0.50\pm0.01$ &  0.042  \\ 
2012 & $0.49\pm0.02$ &  0.040  \\ 
2014A & $0.543\pm0.009$ &  0.030  \\ 
2014B & $0.57\pm0.02$ &  0.047  \\ 
2018 & $0.545\pm0.007$ &  0.052  \\  
2019 & $0.550\pm0.008$ &  0.047  \\ 
\enddata
\tablenotetext{a}{Data from \cite{Jewitt2002}}
\tablenotetext{b}{Data from \cite{Lellouch2002}}
\tablenotetext{c}{Data from \cite{Hicks2005}}
\end{deluxetable}

\section{Varuna's pole orientation and shape}
\label{sec:analysis}

The aspect angle of a solar system body, formed by the line of sight and the rotational axis of the body, changes due to the orbital motion of the body around the Sun. This change produces a variation in the geometric cross section of the object's rotational modulation that is translated into a variation in the amplitude which, for a triaxial ellipsoid, is described as
\begin{equation}
\label{eq:amplitude_axesratio}
\Delta m=-2.5\log\left[\frac{b}{a}\left( \frac{a^2\cos^2(\delta)+c^2\sin^2(\delta)}{b^2\cos^2(\delta)+c^2\sin^2(\delta)}\right)^{1/2}\right],
\end{equation}
where $\delta$ is the aspect angle and $a$, $b$, $c$ are the semi-axes of a triaxial body \citep[with $a>b>c$;][]{Tegler2005}. The aspect angle of the body can be written as
\begin{equation}
\label{eq:aspect_angle}
\delta=\frac{\pi}{2}-\arcsin\left[\sin(\beta_{\rm e})\sin(\beta_{\rm p})+\cos(\beta_{\rm e})\cos(\beta_{\rm p})\cos(\lambda_{\rm e}-\lambda_{\rm p})\right],
\end{equation}
where $\lambda_{\rm e}$ and $\beta_{\rm e}$ are the ecliptic longitude and latitude of Varuna's-centered reference frame (given by the ephemeris), and $\lambda_{\rm p}$ and $\beta_{\rm p}$ are the ecliptic longitude and latitude of the pole orientation \citep{Schroll1976}. These coordinates change with the observing epochs. In order to fit the observational data to the equation (\ref{eq:amplitude_axesratio}), we carried out a grid search for the free parameters in the equations: $b/a$, $c/b$, $\lambda_{\rm p}$ and $\beta_{\rm p}$, which gave theoretical values for $\Delta m$ \citep[as done in][]{FernandezValenzuela2017}. In a first step, we assumed hydrostatic equilibrium, thus reducing the number of free parameters to three, as $c/b$ is then related to $b/a$ \citep{Chandrasekhar1987}. We explored $b/a$ in the range [0.44, 0.60] at intervals of 0.02. The upper limit of 0.60 is imposed by the measurement of the largest observed amplitude up to date (see table \ref{tb:amplitudes}). The lower limit of 0.44 is imposed by theoretical models of rotational equilibrium configurations of strengthless bodies \citep{Leone1984}. We explored the possible values of $\lambda_{\rm p}$ and $\beta_{\rm p}$ on the entire sky at intervals of $0.5^{\circ}$.

The goodness of the fit is evaluated using the $\chi^2_{\rm pdf}$ test. The possible solutions are those that have values of $\chi^2_{\rm pdf}$ within the range $\chi^2_{\rm pdf,min}+0.584$ \cite[from the $\chi^2$ distribution with a 0.9 level of confidence for 3 degrees of freedom;][]{Feller1971}. As a result, the subsequent range of values for the axis ratio $b/a$ is $[0.56,0.60]$, where the lower limit is given by the goodness of the fit. Considering the results from the different axis ratios, we constrained the pole orientation with $\lambda_{\rm p}\in[43.0,83.5]^{\circ}$ and $\beta_{\rm p}\in[-70.0,-58.0]^{\circ}$ and $\lambda_{\rm p}\in[209,220]^{\circ}$ and $\beta_{\rm p}\in[-49.0,-35.5]^{\circ}$ (with their complementary directions also possible).

\begin{figure}
\plottwo{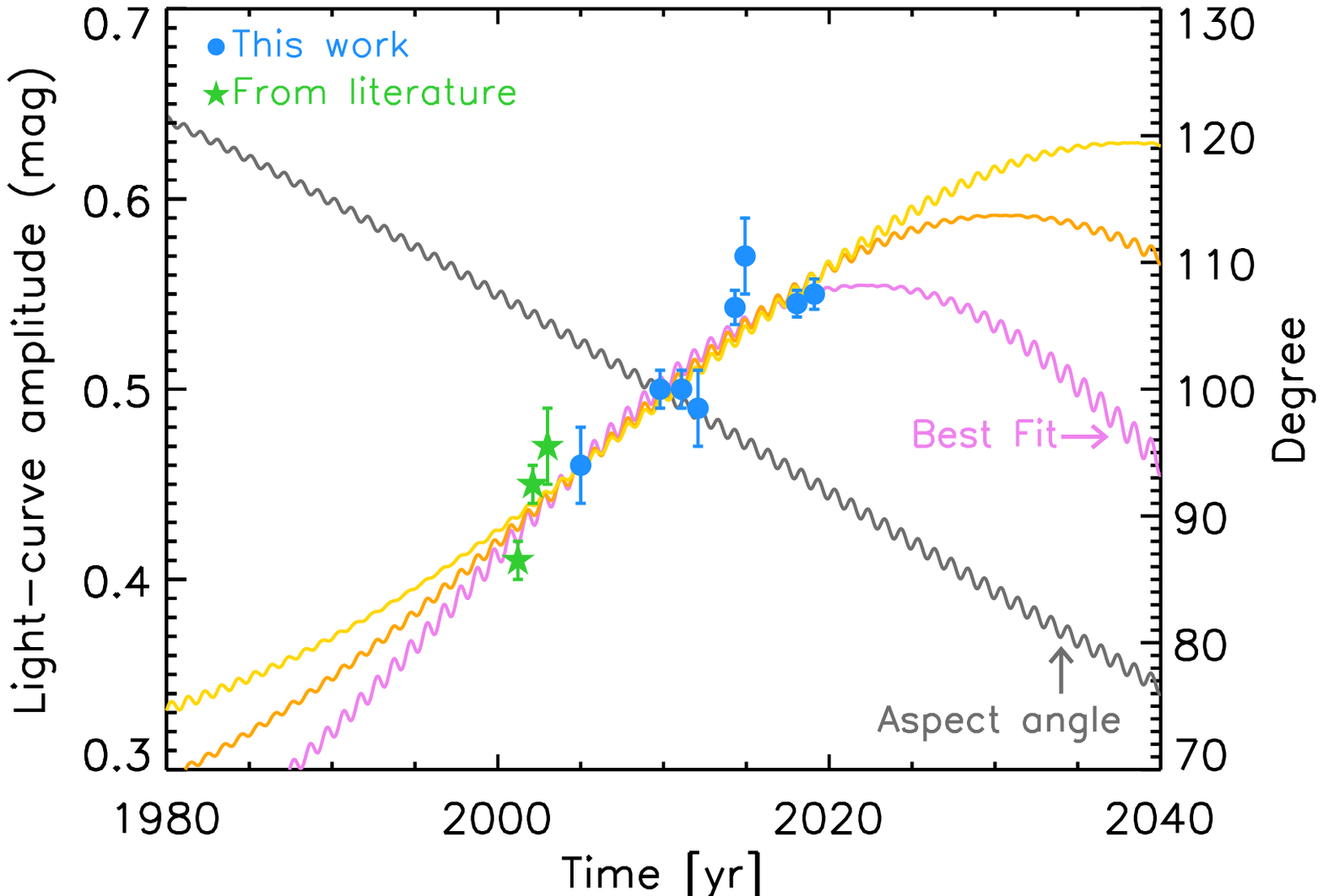}{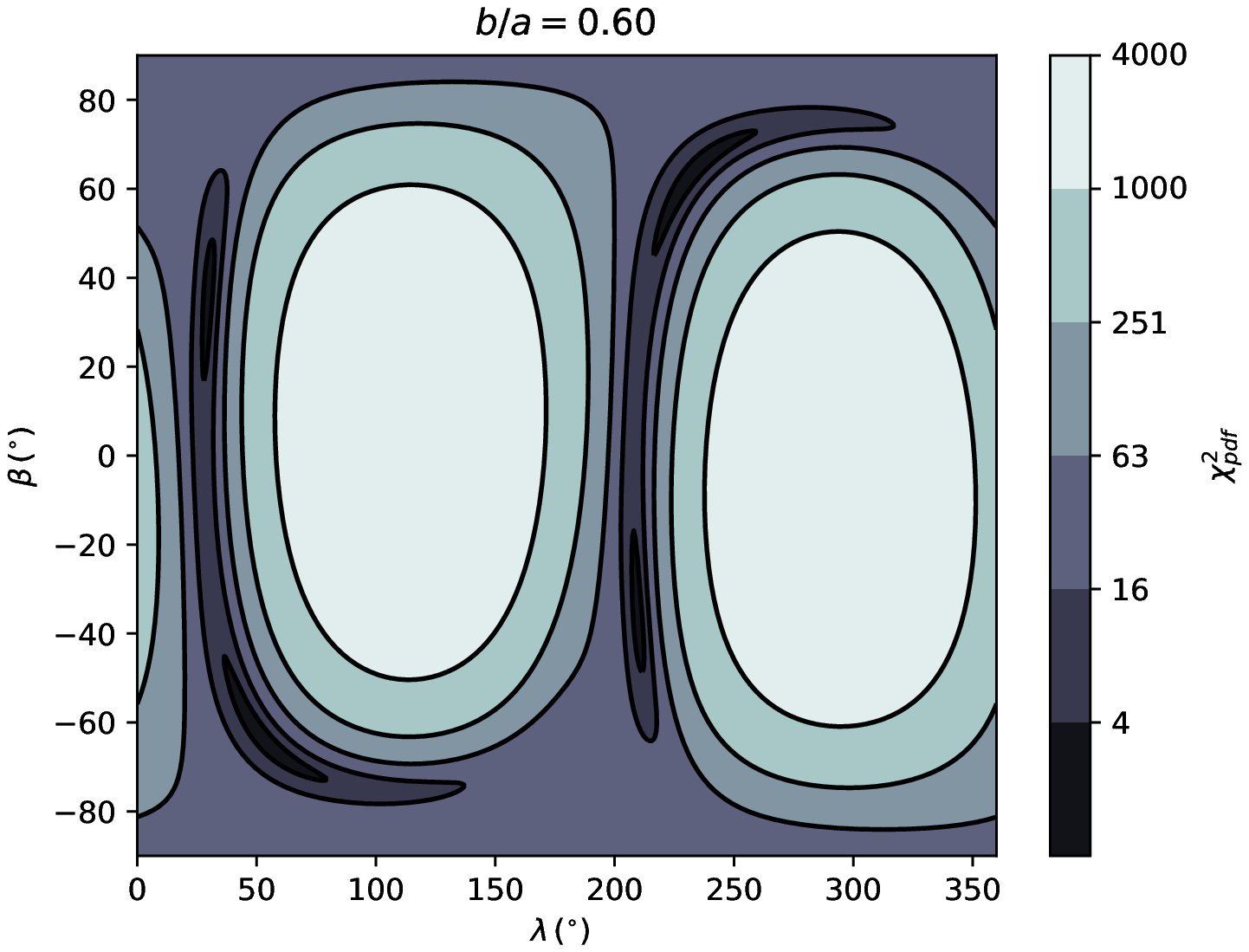}
\caption{Left panel: Model of the amplitude given by the fit to the observational data using $b/a=0.60$ (pink line, best fit), 0.58 (orange line), and 0.56 (yellow line). The grey line represents the Varuna's aspect angle given by the best fit. Right panel: $\chi^2_{\rm pdf}$ map of possible values for $b/a=0.60$. Similar configurations occur for values of the axis ratios of 0.58 and 0.56. \label{fig:pole_orientation}}
\end{figure}

Specifically, we obtained the $\chi^2_{\rm pdf,min}=1.904$ for the following parameters: semi-axis ratio $b/a=0.60$ (with $c/b=0.72$ imposed by the hydrostatic equilibrium), and pole orientation $\lambda_{\rm p}=53^{\circ}\pm10^{\circ}$ and $\beta_{\rm p}=-64^{\circ}\pm6^{\circ}$ (the complementary direction $\lambda_{\rm p}=233^{\circ}$ and $\beta_{\rm p}=64^{\circ}$ is also possible for the same $\chi_{\rm pdf}^2$ value). The errors of $\lambda_{\rm p}$ and $\beta_{\rm p}$ have been obtained considering the values that fit within the interval of $\chi^2_{\rm pdf,min}+0.584$ for this specific axis ratio. Figure \ref{fig:pole_orientation} shows the best fit given by the above values (left panel) and a $\chi^2_{\rm pdf}$ map as a function of $\lambda_{\rm p}$ and $\beta_{\rm p}$ (right panel) for the specific value of $b/a=0.60$. The solution in the range of $\lambda_{\rm p}\in[209,211]^{\circ}$ and $\beta_{\rm p}\in[-42.0,-35.5]^{\circ}$ can be discarded making use of results from the stellar occultation that occurred in 2010 \citep{Sicardy2010}. Even if a detailed analysis will be published elsewhere, we can advance that this yields restricted intervals for the values of $\lambda_{\rm p}$ and $\beta_{\rm p}$. It turns out that the first possibility $\lambda_{\rm p}=53^{\circ}\pm10^{\circ}$, $\beta_{\rm p}=-64^{\circ}\pm6^{\circ}$ (as well as its complementary direction) lies inside these intervals, for which the resulting area of Varuna does not produce a geometric albedo being unnaturally small\footnote{The relation between the equivalent diameter and the geometric albedo is $D=Cp^{-1/2}10^{-H/5}$, where $C=1329$ km is a constant, $p$ is the geometric albedo and $H$ is the absolute magnitude.}, but this is not the case for $\lambda_{\rm p}\in[209,211]^{\circ}$, $\beta_{\rm p}\in[-42.0,-35.5]^{\circ}$, for which the size of Varuna would be too large, corresponding to an albedo well below the lower limit of 0.04 for TNOs.

In a second step, we relaxed the assumption of hydrostatic equilibrium, adding the parameter $c/b$ to the grid search in the range [0.5, 1.0] using steps with a length of 0.1. However, this does not lead to appreciable differences. The smallest value of $\chi^2_{\rm pdf}$ found is 1.905, which results in values of $\lambda_{\rm p}=54^{\circ}$ and $\beta_{\rm p}=-65^{\circ}$, with the axis ratios $b/a = 0.60$ and $c/b=0.70$, which is the equivalent result to the case of hydrostatic equilibrium. Currently, from the best fit, Varuna is close to an ``edge-on'' configuration (see left panel in figure \ref{fig:pole_orientation}), where the aspect angle is close to $90^{\circ}$. Therefore, the c-axis, which is the one that would determine whether Varuna is in hydrostatic equilibrium or not, does not produce a strong effect on the light curve shape, being difficult to distinguish between both scenarios.

There also exists the possibility that Varuna could be a contact binary, as happened with 2014 MU$_{19}$ \citep{Stern2019}. However, this scenario was explored by \cite{Jewitt2002} and was considered very unlikely for Varuna. Note that contact binaries have specific V-shaped light curves \citep{Thirouin2017a,Thirouin2017b}, which is not the case of Varuna.

\section{Indications for Varuna's close-in satellite}
\label{sec:satellite}

\begin{figure}
 \plotone{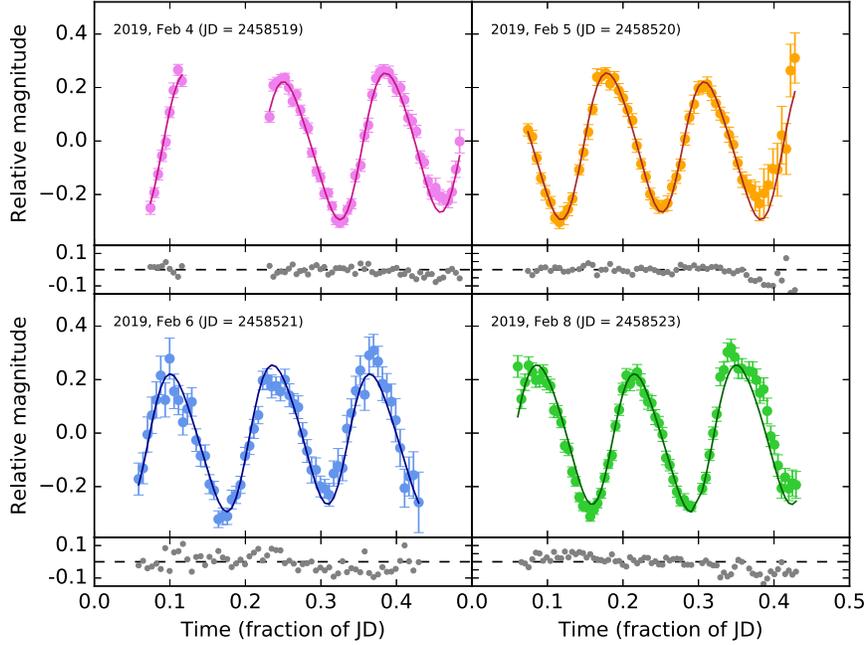}
\caption{Light curves observed during 2019, each panel correspondes to a different day in which the corresponding data are represented by circles. The solid lines over-plotted to the circles represent the fourth-order Fourier fit to the observational data, with each day fitted independently. The $x$-axis represent the fraction of JD, which is given within the corresponding panel. The residuals to the fit can be seen at the bottom panels.  \label{fig:RLC2019}}
\end{figure}

\begin{figure}[ht!]
\plottwo{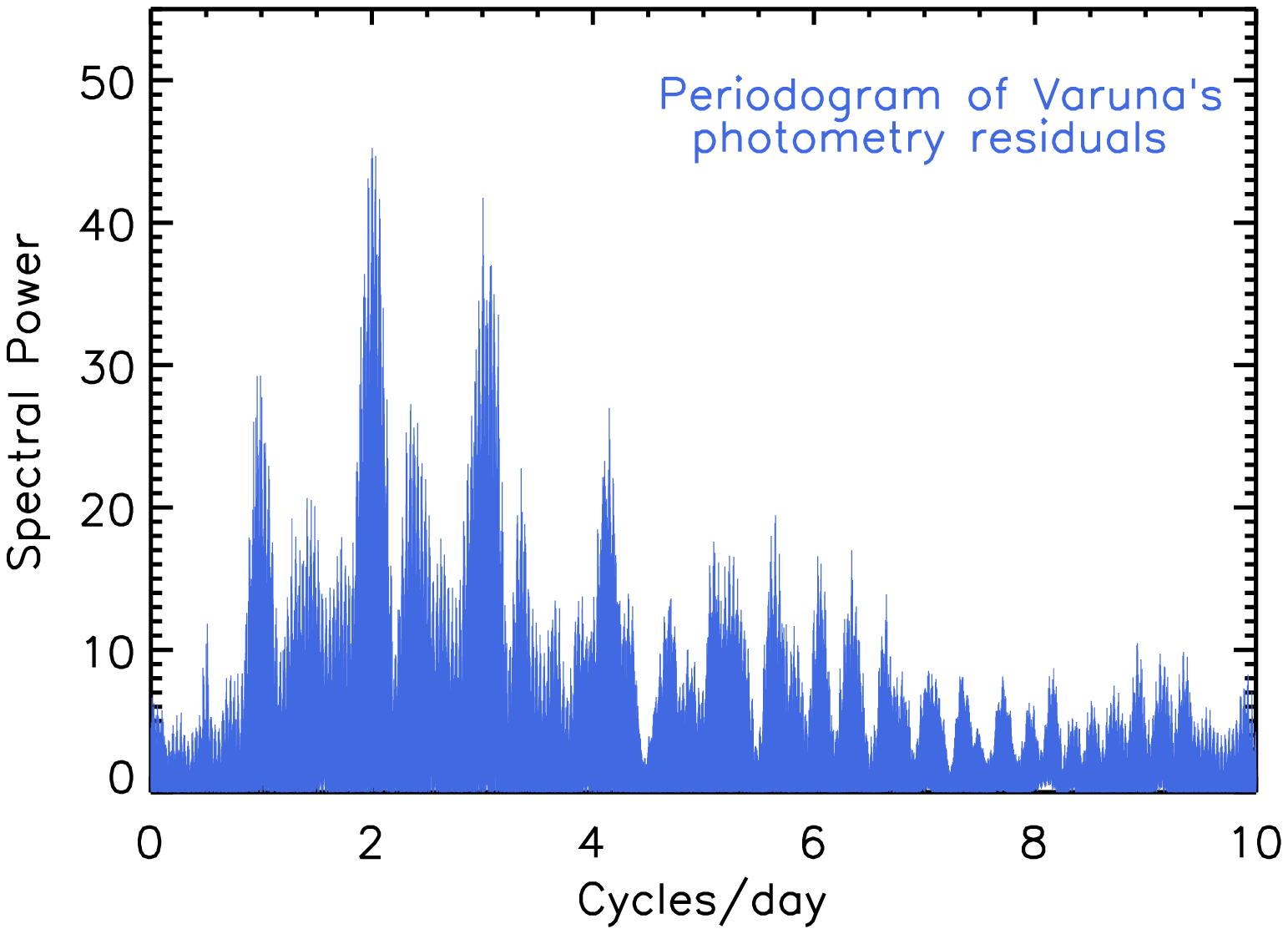}{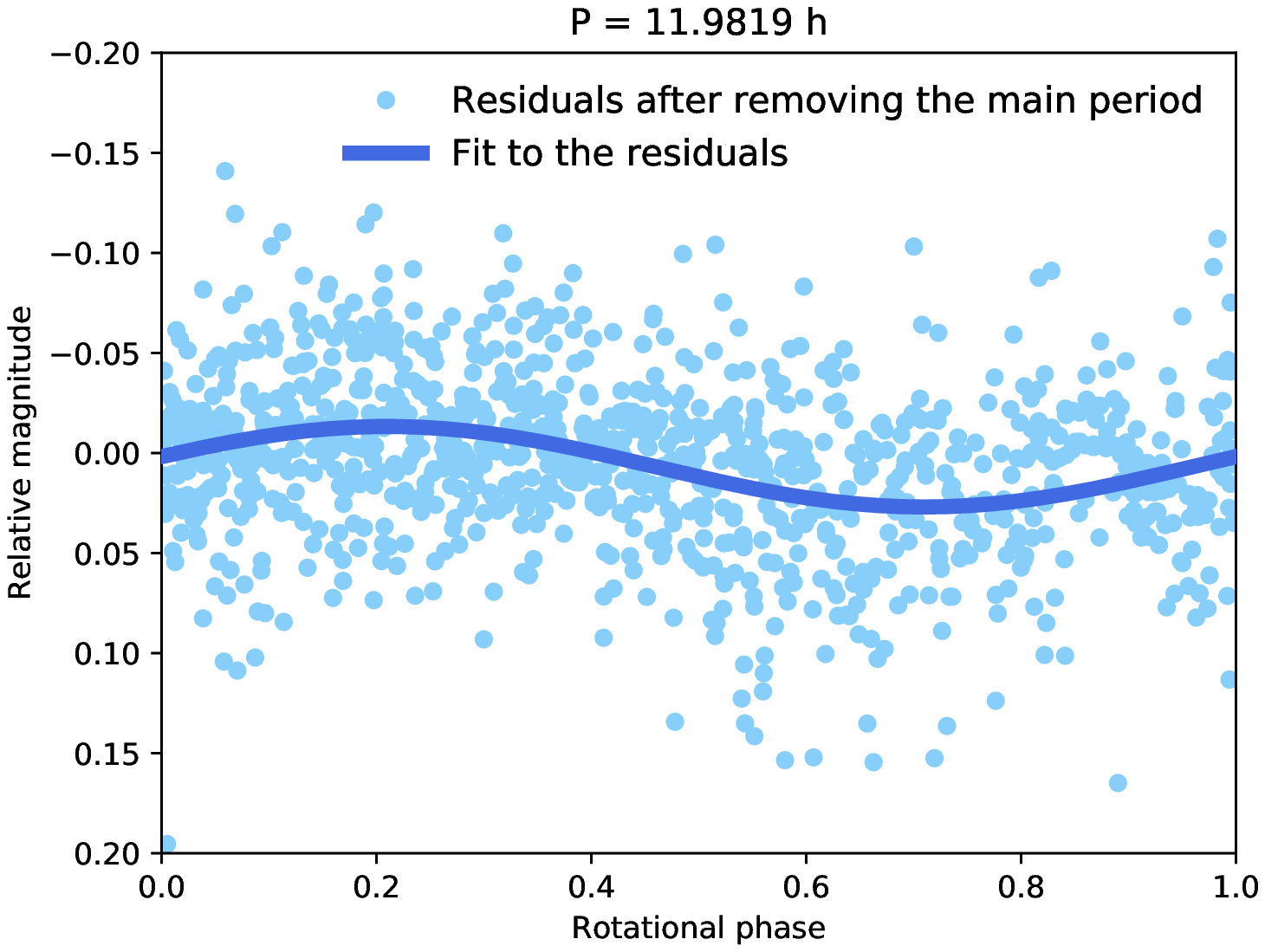}
\caption{Left panel: Periodogram of the residuals of the fit to Varuna's observational data from years 2011, 2014, 2018 and 2019. A maximum spectral power of 45 is obtained at 2.003 cycles/day (11.9819 h). Right panel: Residuals of Varuna's observational data from years 2011, 2014, 2018 and 2019 (blue circles), folded to the 11.9819 h period. The blue line represents a one-order Fourier function fit to the points. \label{fig:periodogram}}
\end{figure}

Even though the signal-to-noise ratio (SNR) of the data in most of the runs was good (well above 50), the residuals of the Fourier fits were considerably scattered, indicating that the model may have been missing some aspect. We focused on the data obtained during 2019, with the best SNR ($\sim80$) for Varuna. In figure \ref{fig:RLC2019}, we plot the relative magnitude of each day versus Julian date. While the four light curves present a good fit, at localized points there are deviations of the model with respect to the data. Moreover, when merging the four days, the quality of the assembled light curve decreases considerably, as the scatter of residuals increases with respect to single night fits and the parameters of the fits change slightly from night to night. A satellite might be responsable for changes in the shape of the main Varuna's light curve. In order to test the potential presence of a rotating satellite that could be generating an additional periodicity in the photometry, we used the residuals from the fit of the 2011, 2014A, 2014B, 2018 and 2019 light curves (which are the ones with higher quality and better sampled) and searched for a period using the Lomb periodogram technique \citep{Lomb1976} as implemented in \cite{Press1992}. This technique has been largely applied to detect satellites in asteroids \cite[e.g.,][]{Pravec1997,Pravec2002,Margot2015}. In figure \ref{fig:periodogram} we plotted the Lomb periodogram of the residuals (left panel), which gives the maximum spectral power of 45 for the frequency of 2.003 cycles/day (11.9819 h), and the residuals folded to that period (right panel), where the blue line shows a one-order Fourier function fit. The amplitude obtained from this fit, 0.04 mag, is related to the rotational modulation of the satellite. Assuming that the satellite's light curve is due to an albedo spot and that its spin and orbit are synchronized (we will check below that this is a reasonable assumption), the distance $d$ at which the satellite orbits from Varuna is given by the equation:
\begin{equation}
d^3=\frac{T_{\rm S}^2GM}{4\pi},
\end{equation}
where $T_{\rm S}$ is the satellite's orbital period (in seconds), $G$ is the gravitational constant and $M=M_{\rm V}+M_{\rm S}$ is the mass of the system. Varuna's mass, $\sim 10^{21}$ kg, is estimated taking into account a density of 1100 kg$\,$m$^{-3}$, assuming hydrostatic equilibrium in a Jacobi ellipsoid with axis ratio $b/a\in[0.60-0.56]$, and volume-equivalent diameter of 700 km. This results in an estimation of $d\sim1300$ km, which places the satellite outside the Roche limit,
\begin{equation}
\label{eq:roche_limit}
d_{\rm R}=a_{\rm V}\left(2\frac{\rho_{\rm V}}{\rho_{\rm S}}\right)^{1/3},
\end{equation}
where $a_{\rm V}\simeq550$ km and $\rho_{\rm V}$ are Varuna's largest semi-axis and density, respectively, and $\rho_{\rm S}$ is the satellite's density. Assuming that the satellite has a density of $\sim300$ kg$\,$m$^{-3}$ \citep[similar to 2014 MU$_{69}$'s density;][]{Stern2019b}, this results in a Roche limit of 1000 km.

 We should note that the satellite could have an elongated shape with a double-peaked light-curve. This implies that the rotation period of the satellite would be double ($\sim24$ h), and so its orbital period. This would locate the satellite $\sim2000$ km from Varuna, with no strong implications for the following calculations. The angular distance between Varuna and the satellite is then smaller than the resolution of the Wide Field Camera 3 installed at the Hubble Space Telescope, thus making imposible to resolve the system.
 
The time required by the satellite to be tidally locked, $t_{\rm lock}$, can be obtained using the equation given by \cite{Hubbard1984},
\begin{equation}
\label{eq:Hubbard}
t_{\rm lock}=\frac{2\pi M_{\rm V}d^6}{3k_{\rm V}GM^2_{\rm S}a^3_{\rm V}T_{\rm V0}\delta},
\end{equation}
where $M_{\rm V}$ is Varuna's mass, $k_{\rm V}$ is the secular Lomb number which has a value of 3/2 for homogeneous bodies, $M_{\rm S}$ is the mass of the satellite, $T_{\rm V 0}$ is the initial rotation rate of Varuna and $\delta$ is expressed as $\arctan(1/Q)$ with $Q$ is the dissipation. Assuming a dissipation of $Q=100$, and that $T_{\rm V 0}$ was greater or equal than $T_{\rm V}$ currently measured, it is straightforward to calculate the dimensionless number $t_{\rm lock}/t_{\rm ss}$ with $t_{\rm ss}\simeq1.5\times 10^{17}\mbox{ s}$ the age of the solar system, obtaining $t_{\rm lock}/t_{\rm ss}\lesssim 5\times10^9\mbox{ kg}\times\left(M_{\rm V}/M_{\rm S}^2\right)$. Using Varuna's largest semi-axis and its density, it follows that, as long as $M_{\rm S}/M_{\rm V}\gg 10^{-6}$, then the dimensionless number $t_{\rm lock}/t_{\rm ss}$, is small. This validates the assumption that the satellite is tidally locked, as the time required by the the satellite to enter this configuration is much smaller than the age of the solar system.

The value obtained for the pole orientation (section \ref{sec:analysis}), places the body almost in an ``edge-on'' configuration, implying that mutual events produced by the satellite may occur in the coming years, assuming that the satellite's orbit is in Varuna's equatorial plane.

\section{Conclusions}
\label{sec:conclusions}

Time-series photometry of Varuna at different epochs have revealed a remarkable change in the amplitude of the light curve of $\sim0.13$ mag. A geometric model that explains this change as the rotational variability due to Varuna's shape provides a range of solutions for its pole orientation of $\lambda_{\rm p}\in[43.0,83.5]^{\circ}$ and $\beta_{\rm p}\in[-70.0,-58.0]^{\circ}$ and its axes ratio $b/a\in[0.56,0.60]$. Nowadays, only around four TNOs have well determined pole orientations (Pluto, Charon, Haumea and 2014 MU$_{69}$), with three of them visited by spacecrafts. Additionally, we noticed that Varuna's light curves have deviations from the expected behavior of a simple ellipsoidal body, and we analyzed the periodogram of the residuals of the photometric fits to the data, finding a peak of high spectral power for a period of 11.9819 h. This appears consistent with the idea of the existence of a close-in satellite, separated from Varuna a distance of $\sim1300$ km (depending on Varuna's mass which is not accurately known). Until now, satellite discoveries in the trans-Neptunian region have been only accomplished using direct imaging \citep{Noll2008}. With this work, we have illustrated that photometric techniques can be used to reveal the existence of a satellite for an object in the trans-Neptunian region. The light-curve technique has the potential to reveal objects much closer-in than the direct imaging technique, helping to remove biases in our current knowledge of the population of binaries or bodies with satellites.

\section*{Acknowlegements}

EFV acknowledges UCF 2017 Preeminent Postdoctoral Program (P$^3$). Part of the research leading to these results has received funding from the European Unions Horizon 2020 Research and Innovation Programme, under Grant Agreement No. 687378 (SBNAF). We would like to acknowledge financial support by the Spanish grant AYA-2017-84637-R and the financial support from the State Agency for Research of the Spanish MCIU through the ``Center of Excellence Severo Ochoa'' award for the Instituto de Astrof\'isica de Andaluc\'ia (SEV- 2017-0709). This work is partially based on observations made with the Italian Telescopio Nazionale Galileo (TNG) operated on the island of La Palma by the Fundaci\'on Galileo Galilei of the INAF (Istituto Nazionale di Astrofisica) at the Spanish Observatorio del Roque de los Muchachos of the Instituto de Astrof\'isica de Canarias. We thank the director of the Telescopio Nazionale Galileo for allocation of Director's Discretionary Time. We acknowledge Ra\'ul Carballo-Rubio for providing useful comments that helped improving this manuscript and Pedro Guti\'errez for helpful discussions.

\newpage

\section*{Online material}

\begin{center}
\begin{table}[h]
\begin{center}
 \renewcommand\thetable{1}
\caption{Log of observations.}
\begin{tabular}{|cccc|}
\hline
Date		&	Telescope	& Seeing ($''$)	& Number of images\\
\hline
2005		& 1.5-m OSN		& 2.01 			&	75	\\

2009		& 1.5-m OSN 		& 1.85			&	48	\\
		& 2.2-m CAHA		& 1.40			&	28	\\

2011		& 1.5-m OSN		& 2.8			&	120	\\
		& 2.2-m CAHA		& 1.8			&	38	\\
		
2012		& 1.5-m OSN		& 2.6			&	64	\\
2014 semester A	& TNG			& 1.6			&	119	\\
2014 semester B & 1.23-m CAHA		& 1.7			&	116	\\
2018		& 1.23-m CAHA		& 1.8			&	135	\\
		& 1.5-m OSN		& 2.1			&	255	\\
2019		& 2.2-m CAHA		& 1.5			&	119	\\
		& 1.5-m OSN		& 3.9			&	64	\\
		& 1.23-m CAHA		& 1.7			&	75	\\	
		\hline
\end{tabular}
\end{center}
\end{table}

 \renewcommand\thetable{2}
\begin{longtable}{|c c c c c c|}

\caption{Results from the relative photometry observations and observational circumstances. Abbreviations are define as follows: Julian date (JD), relative magnitude (Rel. mag.), error of the relative magnitude (err.), heliocentric and topocentric distances ($r_{\rm H}$ and $\Delta$, respectively), and phase angle ($\alpha$).}\\
\hline
\textbf{JD} & \textbf{Rel. mag.} & \textbf{err. } & \textbf{$r_{\rm H}$} & \textbf{$\Delta$} & \textbf{$\alpha$}\\
 & \textbf{(mag)} & \textbf{(mag)} & \textbf{(au)} & \textbf{(au)} & \textbf{(deg)}\\
\hline
\endfirsthead
\multicolumn{4}{c}%
{\tablename\ \thetable\ -- \textit{Continued from previous page}} \\
\hline
\textbf{JD} & \textbf{Rel. mag.} & \textbf{err. } & \textbf{$r_{\rm H}$} & \textbf{$\Delta$} & \textbf{$\alpha$}\\
 & \textbf{(mag)} & \textbf{(mag)} & \textbf{(au)} & \textbf{(au)} & \textbf{(deg)}\\
\hline
\endhead
\hline \multicolumn{4}{r}{\textit{Continued on next page}} \\
\endfoot
\hline
\endlastfoot

2453376.23	&	0.1137	&	0.023	&	43.2515	&	42.3639	&	0.5659	\\
2453376.236	&	0.061	&	0.0225	&	43.2515	&	42.3639	&	0.5659	\\
2453376.242	&	0.0272	&	0.021	&	43.2515	&	42.3639	&	0.5659	\\
2453376.248	&	-0.0367	&	0.0197	&	43.2515	&	42.3639	&	0.5659	\\
2453376.254	&	-0.1125	&	0.0205	&	43.2515	&	42.3639	&	0.5659	\\
2453376.26	&	-0.1196	&	0.0195	&	43.2515	&	42.3639	&	0.5659	\\
2453376.266	&	-0.1338	&	0.0201	&	43.2515	&	42.3639	&	0.5659	\\
2453376.272	&	-0.1345	&	0.0192	&	43.2515	&	42.3639	&	0.5659	\\
\end{longtable}
\end{center}



\end{document}